# Automatic Generation of OWL Ontology from XML Data Source

**Nora Yahia [1], Sahar A. Mokhtar [2] and AbdelWahab Ahmed [3]**

**[1, 2] Computers and Systems Department, Electronic Research Institute**
**Dokki, 12622/ Cairo, Egypt**
nora@eri.sci.eg
sahar@eri.sci.eg

**[3] Computers and Systems Engineering Department, Faculty of Engineering, Al-Azhar University**
**Nasr City, 11371 / Cairo, Egypt**
*wahab2410@hotmail.com*

**Abstract**

The eXtensible Markup Language (XML) can be used as data exchange format in different domains. It allows different parties to exchange data by providing common understanding of the basic concepts in the domain. XML covers the syntactic level, but lacks support for reasoning. Ontology can provide a semantic representation of domain knowledge which supports efficient reasoning and expressive power. One of the most popular ontology languages is the Web Ontology Language (OWL). It can represent domain knowledge using classes, properties, axioms and instances for the use in a distributed environment such as the World Wide Web. This paper presents a new method for automatic generation of OWL ontology from XML data sources.

## 1. Introduction

Information sources can be structured such as relational databases, or semi-structured such as XML data sources [1]. XML has reached consensus among most standards bodies, becoming the de facto standard format for data exchange. Even though XML has only been launched in recent years and is still undergoing rapid development, nevertheless, XML becomes adopted widely in information management [2]. It has been seen as the most powerful advance the Internet has seen for years. It provides a format that is human readable and machine interpretable. Its simplicity and suppleness of usage fits well with an underlying technology that gives "portability" of information across platforms, applications, and organizations, which is vital for information exchange requirements but it lacks support for expressing semantics of domain knowledge. Formal conceptual models, or ontologies, are necessary to express the semantics of the data. Data sources can be heterogeneous in syntax, schema, or semantics, thus making data interoperation a difficult task [3]. Syntactic heterogeneity is caused by the use of different models or languages. Schematic heterogeneity results from structural differences. Semantic heterogeneity is caused by different meanings or interpretations of data in various contexts. Ontologies provide a promised technology to solve the semantic heterogeneity problem, because they allow to explicitly representing common semantics of a domain of discourse. An ontology can be defined as a formal, explicit specification of a shared conceptualization [4]. 'Conceptualization' refers to an abstract model of phenomena in the world by having identified the relevant concepts of those phenomena. 'Explicit' means that the type of concepts used, and the constraints on their use are explicitly defined. 'Formal' refers to the fact that the ontology should be machine readable. 'Shared' reflects that ontology should capture consensual knowledge accepted by the communities. An ontology formally defines different concepts of a domain and relationships between these concepts. In ontology-based approaches for information integration, local ontologies are used to describe the semantics of local information sources. The more recent Web Ontology Language (OWL) [5], has become a popular standard for data representation and exchange. The OWL supports the representation of domain knowledge using classes, properties and instances for the use in a distributed environment as the World Wide Web. This paper proposes a new strategy for generating OWL ontologies automatically out of existing XML data sources. This has to be done by establishing suitable mappings between the different data model elements of XML and OWL. OWL includes three sub languages called:

- OWL Lite: it supports those users who primarily need a classification hierarchy and simple constraints .
- OWL (Description Logic) DL: it supports those users who want the maximum expressiveness while retaining computational completeness (all conclusions are guaranteed to be computed) and decidability (all computations will finish in finite time).

- OWL Full: it used by users who want maximum expressiveness and the syntactic freedom of Resource Description Framework (RDF) with no computational guarantees.

There is a tradeoff between expressive power and efficient reasoning support. The richer the language is, the more inefficient the reasoning support becomes. OWL-DL provides a compromise: a language supported by reasonably efficient reasoners and a language that can express large classes of ontologies and knowledge. For this reason the proposed generated ontology is encoded in OWL-DL format.

This paper is organized as follows: Section 2 presents several strategies for deriving ontologies from heterogeneous XML data sources. Some strategies targeted either more on a general mapping between XML and RDF others aim at mapping XML Schema to OWL without considering XML instance data. Section 3 presents the proposed approach used to mapping XML data source to OWL ontology. Section 4 focuses on the experimental and results. Section 5 presents the comparative and analysis. Section 6 contains the conclusion and future work.

## 2. Related work

In this section, several strategies for deriving ontologies from heterogeneous XML data sources have been developed. Some approaches targeted either more on a general mapping between XML and RDF others aim at mapping XML Schema to OWL. OWL and RDF are much of the same thing, but OWL is a stronger language with greater machine interpretability than RDF. OWL comes with a larger vocabulary and stronger syntax than RDF. This section concentrates on the strategies for OWL ontology generation. The following strategies are named according to their authors.

### 2.1 M. Ferdinand [6]

Ferdinand proposed direct mappings from XML Schema to OWL as well as they described mappings from XML to RDF graphs, but these mappings are independent of each other, i.e., the generated instances do not necessarily respect the ontology created from the XML Schema. The XML schema to OWL mapping process is based on a set of interpretation and transformation rules from XML Schema to OWL. XML Schema complexType, Model group definitions and attribute group definitions are mapped to OWL classes. Element and attribute declaration is mapped to an OWL property. More precisely, elements of simpleType and all attributes are mapped to an owl:DatatypeProperty; elements of complexType are mapped to an owl:ObjectProperty. Two types of inheritance are supported by XML Schema: inheritance by restriction and inheritance by extension. Both are mapped to the only inheritance mechanism in OWL: rdfs:subClassOf. The mappings proposed in this approach are intended to be applied for the engineering of web applications. In particular, for enhancing traditional XML languages and tools by the capabilities of OWL reasoners. This approach does not tackle the question how to create the OWL model, if no XML Schema is available.

### 2.2 H. Bohring and S. Auer [7]

Bohring and Auer have developed a tool, called xml2owl, to create an OWL ontology from an XML schema, and convert XML data to OWL instances compliant to the created ontology. Technically they have developed four Extensible Style Sheet Language Transformation (XSLT) [8] instances to transform XML files to OWL, without any other intervention on semantics and structures during the transformation. Finally the OWL file (read ontology) is automatically generated. This method has been also applied to the Ontowiki platform [9]. Bohring and Ferdinand approaches are similar that it is based on a set of transformation rules from XML Schema to OWL. They proposed that named XML Schema Definition (XSD) complex Types with complex content (i.e., a combination of attributes and sequence of elements) are transformed to OWL classes. When an element contains another element, an OWL object property is created between their corresponding OWL classes. OWL datatype properties emerge from XML attributes and from element containing only a literal and no attributes. XML Schema arity constraints like minOccurs or maxOccurs, are mapped to the equivalent cardinality constraints in OWL, minCardinality and maxCardinality. During the ontology creation from an XML schema, the user has no control on the process. That is, the user has no control on the newly created ontology which captures the implicit semantics exist in the XML Schema structure. Therefore the created ontologies are quite primitive, that is, they are not really semantically richer than the mapped XML Schemas. Furthermore they not mention how to apply it on multiple source documents.

Both of Ferdinand's and Bohring's approaches introduce a good basis of rules to create OWL ontologies from XML Schema. However, they address only simple cases and do not refer to complex cases that arise from the reuse of global types and elements. Also, they do not mention how to specify mappings between multiple XML data sources and generated OWL ontology.

### 2.3 R. Ghawi and N.Cullot [10]

R. Ghawi and N. Cullot proposed a tool, called X2OWL, that aims at building an OWL ontology from an XML data source. This method is based on XML schema to automatically generate the ontology structure, as well as, a set of mapping bridges between the entities of the XML data source and the created ontology, mapping bridges contribute into query translation between OWL and XML. This approach addresses simple cases and complex cases that arise from the reuse of global types and elements that are used to create XML schema. XML schemas can be modeled using different styles, some of them use a single global element (root element), and others use multiple global elements, some styles use global types, and others use only local types. However, the mapping method should cope with all possible design patterns [11]. The presented method also includes a refinement step that allows cleaning the mapping bridges and to re-structure the generated ontology, Although the new proposed approach will be take into consideration how to build ontology from multiple XML data sources.

### 2.4 I. Bedini and B. Nguyen [12]

I. Bedini, and N. Benjamin proposed a framework called Janus that aims to generate an ontology from a large source corpus of XML Schemas, They present a set of patterns that enable the direct, automatic transformation from XML Schema into OWL allowing the integration of much XML data in the Semantic Web. They focus on an advanced logical representation of XML Schema components and present an implementation that it is possible to mine XML Schema sources to extract enough knowledge to build semantically correct ontologies with considerable expressivity. Their implementation was shown foremost to be more complete than others, and in particular greatly improves on the number of complex transformation patterns with respect Bohring et al approach. This approach does not tackle the question of how to create the OWL ontology, if no XML Schema is available.

### 2.5 J. Xu and W. Li [13]

Xu and Li proposed an approach to construct OWL ontology from XML document with the help of entity relation model. That is, they propose an XML-to-Relational (XTR) mapping approach to map an XML document to an entity-relation model, and then a Relational-to-Ontology (RTO) mapping approach to map an entity-relation model to an OWL ontology. However, the OWL ontology is expressed using ad-hoc vocabularies for describing relational database, therefore it can not be considered as domain ontology.

The next section presents a new method for automatic generation of OWL ontology from multiple XML data sources.

## 3. The Proposed Method

The new proposed method uses the same notations used in [10] with some modifications to apply on multiple xml data sources. The new proposed approach is based on XML schema to build the ontology. If the schema does not exist, it can be automatically generated from the source XML document, this method copes with all possible complex cases arising from different XML schema design styles.

### 3.1 Design of the Automatic Ontology Building System

As shown in Fig. 1, the generation of OWL ontology from XML data sources could be described in 4 steps which are:

1) The XML document is transformed to XML-Schema using the Trang API for java [14]. The Trang takes as input a schema written in XML syntax and produces as output a schema written in XML-Schema.
2) The XML-Schema is analyzed using XML-Schema Object Model (XSOM) [15]. XSOM is a Java library that allows applications to easily parse XML Schema documents and inspect information in them. It is expected to be useful for applications that take XML Schema as an input.
3) The output of XSOM is used as input for the Java Universal Network/Graph framework (JUNG) [16]. The JUNG is used for graph-based manipulations. It generates XML- Schema Graph (XSG) that describes the schema in the same way whatever its design style is. An XSG is composed of a vertex set, and an edge set. The vertex set contains all elements, attributes, nonprimitive types, element groups and attribute groups. The edge set contains the edges established:
    - From each element to its type (if not primitive),
    - From each type, element group or attribute group to their contained elements and/or attributes.
4) The Jena API [17] uses XSG as input to generate OWL entities. Basically, OWL Classes emerge from complex types, element group declarations, and attribute-group declarations according to the mapping rules which will be discussed in section 3.2. Object properties emerge from element-subelement relationships. Datatype properties emerge from attributes and from simple types.

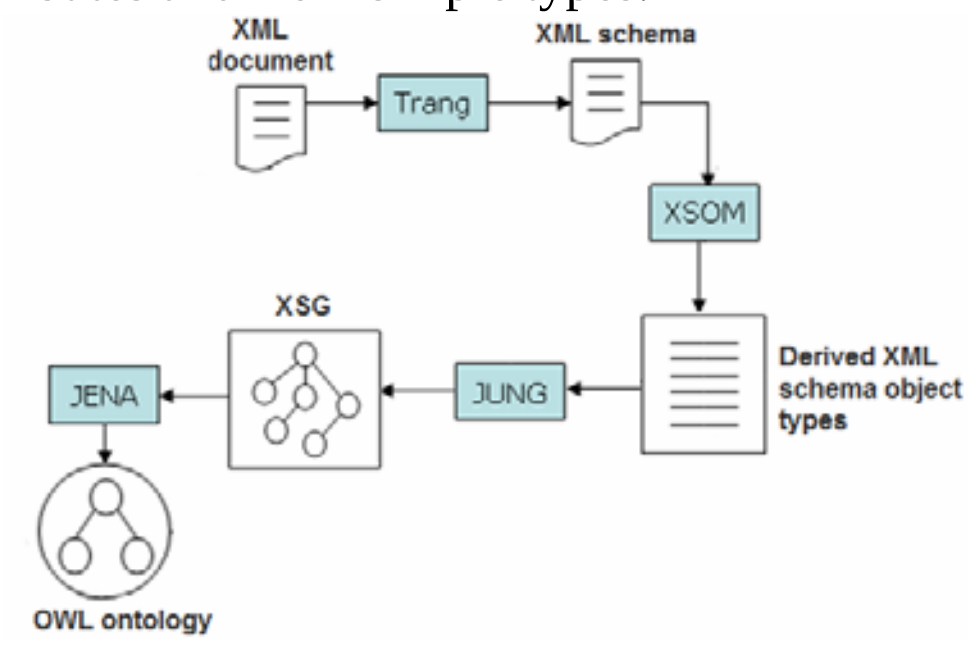

Fig. 1 The generation process of OWL ontology from each XML data source

### 3.2 The generated OWL entities construction

This section presents how to specify XML-to-OWL mappings, the new proposed method is based on mapping rules of [6], [7], and [10], Three types of mappings are presented as follows:

1) **OWL classes**: two kinds of complex types are distinguished: 1) global, named complex types, and 2) local anonymous complex types. Both cases are mapped to the proposed OWL classes. However, a class generated from a global named type will have the name of that type, while a class generated from local anonymous type will have the name of the (only) surrounding element. Element-group and attribute group declarations are also mapped to the proposed OWL classes. XML schema supports two mechanisms of inheritance: extension and restriction. Both of these inheritance mechanisms are translated to the class inheritance mechanism of OWL (using rdfs:subClassOf). When a complex type is defined as an extension or a restriction of another base complex type, then the class corresponding to this type is set as subclass of the class corresponding to the base type.
2) **Object properties**: Elements (global or local) are not mapped directly to the ontology, but the element-subelement relationship in the schema is translated as the proposed object property in the ontology. That is, when an element has a complex type, then that complex type is already mapped to an OWL class. Therefore, an object property is added to the ontology having as domain the class corresponding to the surrounding complex type, and having as range the OWL class corresponding to the type of the element. The name of this proposed object property is the concatenation of “has” with the name of range class.
3) **Datatype properties**: Elements of simple types are mapped to the proposed datatype properties. When a complex type (global or local) contains an element of a simple type (primitive or defined) having as domain the class corresponding to the complex type. If the simple type is a primitive XML Schema Definition (XSD) datatype (xsd:string, xsd:integer, ….) then the range of the proposed datatype property is this datatype. Otherwise, if the simple type is defined in the schema, then the range of the proposed datatype property will be xsd:anyType. Attributes are treated as simple elements and will be mapped to the proposed datatype properties. If a complex type is mixed, then the elements that have this type contain text as well as subelements and/or attributes. To take this text into account, a proposed datatype property is added to the ontology having as domain the class corresponding to the surrounding complex type, and having as range “xsd:string” datatype.

The next section introduces the experimental and results of applying the proposed ontology generation process on scientific publications domain.

## 4. Experimental and Results

In order to illustrate the proposed ontology generation process, The XML document from [18] is used as an example, it represents metadata about scientific publications. Part of the XML document is shown in Fig. 2. In this example, the XML root element <bibliography> contains a repeated sequence of bibliographic entry records; each biblioentry record contains the id attribute, author, title and publisher sub elements. The author element has firstname and lastname sub elements. The publisher element has publishername sub element.

```
<bibliography id="personal_identity">
  <biblioentry id="FHIW13C-1234">
    <author>
      <firstname>Godfrey</firstname>
      <surname>Vesey</surname>
    </author>
    <title>Personal Identity: A Philosophical Analysis</title>
    <publisher>
      <publishername>Cornell University Press</publishername>
    </publisher>
    <pubdate>1977</pubdate>
  </biblioentry>
</bibliography>
```

Fig. 2 Part of the XML document used as an example.

As stated in [19], ontologies and XML schemata serve very different purposes. Ontology languages are a means to specify domain theories based on logical representation and XML schemata are a means to provide integrity constraints for information sources (i.e., documents and/or semi-structured data). It is therefore not surprising to encounter differences when comparing XML schema with ontology languages. However, XML schema and OWL ontologies have one main goal in common: both provide vocabulary and structure for describing information about data. As discussed in section 3.1, the second step of ontology building process is that extracts an XML Schema out of the XML instance data. The generated XML-Schema looks like the following as shown in Fig. 3. It corresponds to having all of the elements defined within the global namespace and then referencing the elements.

```
<xs:element name="bibliography">
  <xs:complexType>
    <xs:sequence>
      <xs:element maxOccurs="unbounded" ref="biblioentry"/>
    </xs:sequence>
    <xs:attribute name="id" use="required" type="xs:NCName"/>
  </xs:complexType>
</xs:element>
<xs:element name="biblioentry">
  <xs:complexType>
    <xs:sequence>
      <xs:element maxOccurs="unbounded" ref="author"/>
      <xs:element ref="title"/>
      <xs:element ref="publisher"/>
      <xs:element ref="pubdate"/>
    </xs:sequence>
    <xs:attribute name="id" use="required" type="xs:NCName"/>
  </xs:complexType>
</xs:element>
<xs:element name="author">
  <xs:complexType>
    <xs:sequence>
      <xs:element ref="firstname"/>
      <xs:element minOccurs="0" ref="othername"/>
      <xs:element ref="surname"/>
    </xs:sequence>
  </xs:complexType>
</xs:element>
<xs:element name="firstname" type="xs:NCName"/>
<xs:element name="othername" type="xs:NCName"/>
<xs:element name="surname" type="xs:NCName"/>
<xs:element name="title" type="xs:string"/>
<xs:element name="publisher">
  <xs:complexType>
    <xs:sequence>
      <xs:element ref="publishername"/>
    </xs:sequence>
  </xs:complexType>
</xs:element>
<xs:element name="publishername" type="xs:string"/>
<xs:element name="pubdate" type="xs:integer"/>
</xs:schema>
```

Fig. 3 The generated XML-Schema.

As discussed in the third step of ontology building process in section 3.1, in order to insure that the generated ontology is independent of any specific XML schema design pattern, ontology building process is based on an XSG that describes the schema in the same way whatever its design style is. Fig.4 shows the XSG for the previous XML-Schema. An XSG is a directed acyclic graph (DAG) that has always a unique root vertex which is the vertex of the root element of XML document. An XSG becomes a tree when elements and types declarations are not re-used within the schema.

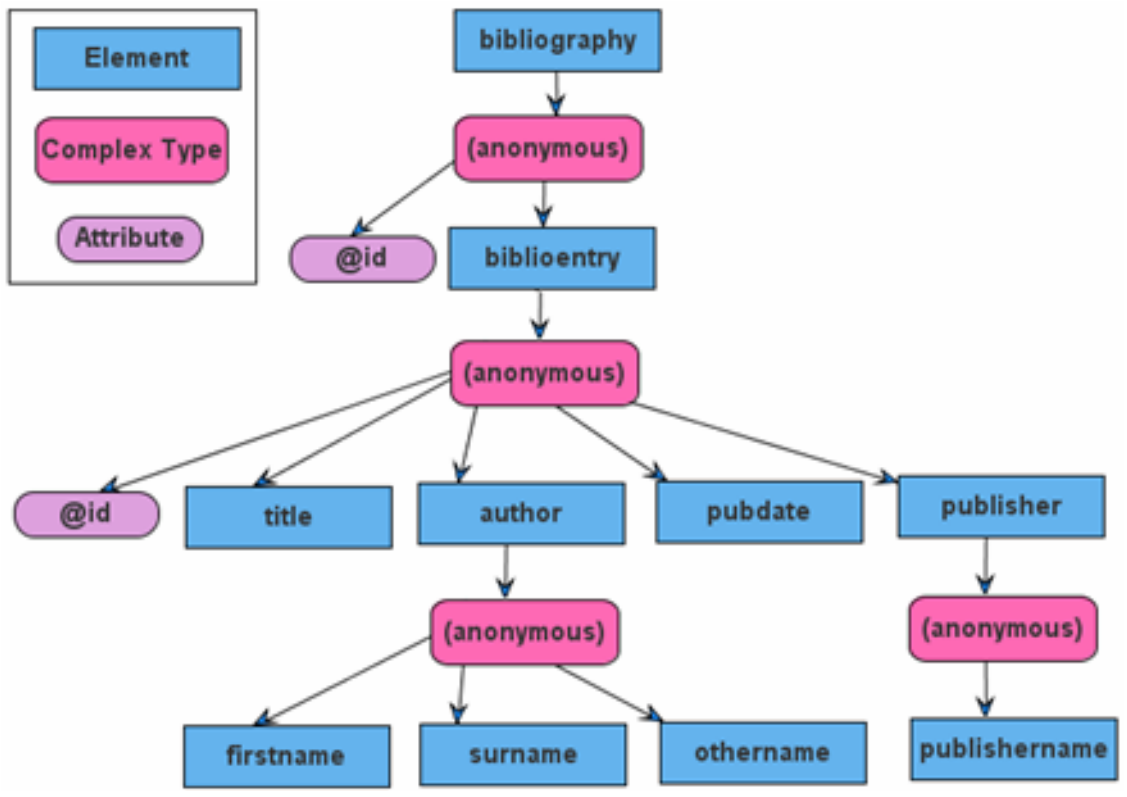

Fig. 4 The generated XSG.

The generated OWL ontology is shown in Fig. 5. In this ontology, there are four locals complex types defined inside the elements bibliography, biblioentry, author and publisher. Therefore, for these complex types, four OWL classes, named: bibliography, biblioentry, author and publisher are created.

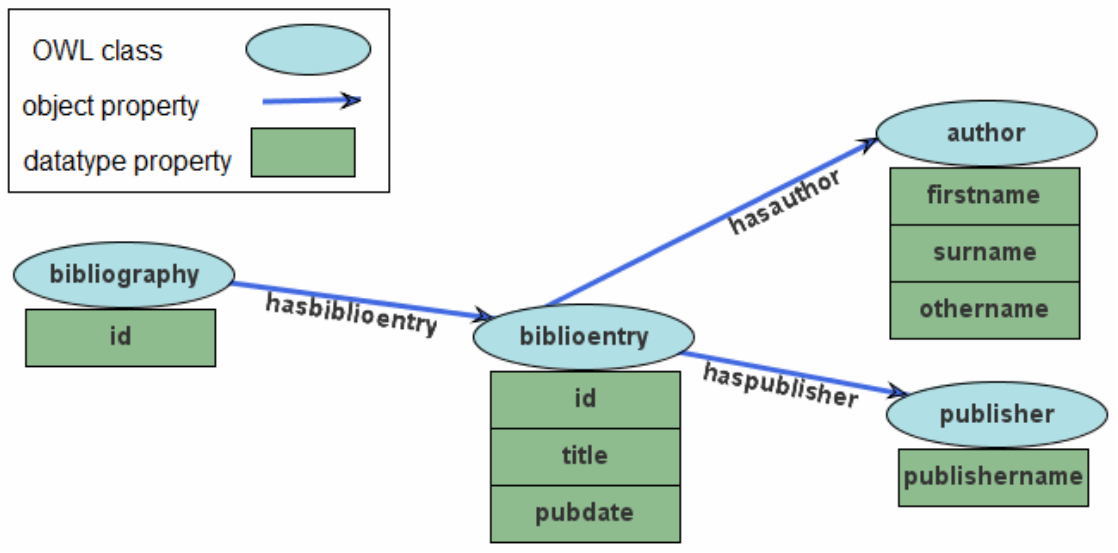

Fig. 5 The generated OWL Ontology.

Tables 1 and 2 show the object properties and datatype properties of the generated ontology in order.

Table 1: The Object properties

| **Object Property** | **Domain** | **Range** |
|---|---|---|
| hasbiblioentry | bibliography | biblioentry |
| hasauthor | biblioentry | author |
| haspublisher | biblioentry | publisher |

Table 2: The Datatype properties

| **Datatype Property** | **Domain** | **Range** |
|---|---|---|
| id | bibliography | NCName |
| id | biblioentry | NCName |
| title | biblioentry | string |
| pubdate | biblioentry | integer |
| firstname | author | NCName |
| surname | author | NCName |
| othername | author | NCName |
| publishername | publisher | string |

## 5. Comparative analysis

As discussed in section 2, there are several approaches for automatic ontology generation process from XML/XML schema, most of cited approaches aimed at creating an ontology from single XML document. The only exception is the proposed approach, in which ontologies are created from heterogeneous XML data sources. I. Bedini and B. Nguyen approach targeted on creating an ontology from merged heterogeneous XML Schemas. The majority of approaches (M. Ferdinand, H. Bohring and S. Auer, R. Ghawi and N. Cullot, I. Bedini and B. Nguyen, and the proposed approach) are based on mapping rules that directly construct ontology from its corresponding XML Schema entities. The approach of J. Xu and W. Li is slightly different because ontology generation process from XML document is based on entity relation model. It is obvious that there are some approaches that addressed simple cases and complex cases that arise from the reuse of global types and elements that are used to create XML schema process, for example R. Ghawi and N. Cullot, I. Bedini and B. Nguyen, and the proposed approach. Other approaches such as: M. Ferdinand, and H. Bohring and S. Auer addressed only simple cases. Concerning the used ontology language, almost all approaches used OWL language. The comparative analysis of methods is summarized in table 3.

Table 3: The Comparative analysis of methods

| | **Input XML Document** | **XML Schema Description** | **Mapping Method** | **Ontology Language** |
|---|---|---|---|---|
| **M. Ferdinand** | Single XML document | XML Schema contains only simple cases | Based on XML Schema | RDF graphs, OWL |
| **H. Bohring and S. Auer** | Single XML document | XML Schema contains only simple cases | Based on XML Schema using XSLT files | OWL |
| **R. Ghawi and N. Cullot** | Single XML document | XML Schema contains simple and complex cases | Based on XML Schema and XSG using mapping bridges and Xpath files | OWL |
| **I. Bedini and B. Nguyen** | XML document not available | Merged heterogeneous XML Schemas contain simple and complex cases | Surjective mapping based on XML mining techniques | OWL |
| **J. Xu and W. Li** | Single XML document | XML Schema not available | Based on entity-relation model | OWL |
| **The proposed approach** | Heterogeneous XML documents | Heterogeneous XML Schemas contains simple and complex cases | Based on XML Schema and XSG | OWL |

## 6. Conclusions and future work

This paper presented a method to automatically generate an OWL ontology from an XML data source. This method is based on XML schema to automatically generate the ontology structure. We have developed a tool as an implementation of the proposed approach. This tool is written in Java and it uses several online-available APIs such as, Jena, Trang, XSOM and JUNG.

Because of the scattered nature of the Web, Information sources can be scattered in different XML data sources. Each information source should be mapped to its own local ontology. The existence of an integrated framework for storing, querying and managing distributed XML data sources is of great importance. For this reason, the subsequent work will be focused on the process of merging the generated local ontologies to generate a global ontology covering the domain knowledge presented in XML data sources.

**Nora Yahia** is an Assistant Research in computers and systems department. She has been working at Electronic Research Institute, Egypt, since 2006. Her fields of research are artificial intelligence, web mining, database, programming language C, Java, C++, and matlab. Nora received her B.Sc. at 2005 from Computer and control Dept., Faculty of Engineering, Zagazig Univ.

**Sahar A. Mokhtar** is an Assistant Prof. She joined the Electronics Research Institute, Egypt, in 1991. Her fields of research are artificial intelligence, machine learning, database, and data mining. She is teaching also database, artificial intelligence and expert systems at Akbar El Yom Academy and Misr University of science and technology. Sahar received her Ph.D. in 2004 from Electronics & Communications Dept., Faculty of Engineering, Cairo Univ. And her M.Sc. at 1994 from Electronics & Communications Dept., Faculty of Engineering, Cairo Univ.

**AbdelWahab Ahmed** is an Associate Professor in Computers and Systems Engineering Department in Al Azhar University. He has worked as a teacher, He is teaching C Language, C++, Visual Basic, Computer Organization, Computer Basics, Matlab, and Computer Engineering. He has worked in many faculties.